\documentclass[12pt]{article}
\newtheorem{theorem}{Theorem}

\let\at=@
\usepackage{latexsym}
\newcommand{\dual}[1]{{\raisebox{0.5pt}{$\stackrel{*}{#1}$}}}
\newcommand{\sfrac}[2]{\textstyle\frac{#1}{#2}}
\begin{document}
\title{PURELY MAGNETIC SPACETIMES}
\author{Barry M Haddow\thanks{{\sl Email:}haddow\at maths.tcd.ie}
\\Department of Mathematics \\
Trinity College\\ Dublin 2\\ Ireland}
\date{}
\maketitle
\begin{abstract}

Purely magnetic
spacetimes, in which the  Riemann tensor
satisfies  $R_{abcd}u^bu^d=0$ for some unit timelike vector
$u^a$, are studied. The algebraic consequences for the  Weyl
and Ricci tensors
are examined in detail and consideration given to the
uniqueness of $u^a$.
Some remarks concerning the
nature of the congruence associated with $u^a$ are made.
\end{abstract}

\newpage

\section{Introduction}

Let $(M,g)$ be a spacetime, that is $M$ is a smooth, Hausdorff, paracompact,
connected 4-dimensional manifold equipped with smooth metric $g$ of signature
+2.
There has been some recent interest  \cite{mcintosh,arianrhod} in spacetimes
whose Riemann tensor \(R^a{}_{bcd}\) satisfies
(Latin indices take the values 1 to 4 throughout)
\begin{equation}
 R_{abcd}u^bu^d =0\label{eq.mag}
\end{equation}
 for some global smooth
 unit timelike vector field $u^a$. Such spacetimes are known
as {\sl purely magnetic}. One can also define a spacetime
to be {\sl purely Weyl magnetic} if the Weyl tensor satisfies a condition
analogous to (\ref{eq.mag})  and clearly the two preceding definitions are
equivalent in vacuum. In  this paper  some
further remarks regarding the existence and uniqueness of solutions
to (\ref{eq.mag}), for $u^a$,  will be made and  the relationship between
purely
magnetic and purely Weyl magnetic spacetimes clarified. In
the vacuum case, some information regarding the rotation and shear
of the timelike congruence associated  with a solution of  (\ref{eq.mag})
will be obtained.

It should also be remarked that condition (\ref{eq.mag})
expresses the fact that \(u^a\) is a {\sl non-generic}
vector field as studied recently by Beem and Harris \cite{beem_h},
who relate  the existence of such vector fields to
properties of the sectional curvature.
The non-existence of non-generic vector fields is an assumption
of some of the singularity theorems \cite{beem_e,h_e}, in that
it is required that if $v^b$ is a tangent vector to a
timelike geodesic then it satisfies
$R_{abcd}v^bv^d\neq 0$ at some point on the geodesic.
If a solution $u^a$
of (\ref{eq.mag}) is tangent to a geodesic then this geodesic
can have no conjugate points, as a simple application of the
Jacobi equation will show.

A related concept is that of a {\sl purely electric spacetime}
in which the Riemann tensor satisfies
\begin{equation}
{}^*R_{abcd}u^bu^d=0\label{eq.el}
\end{equation}
The asterisk denotes the usual dual operation, and
is placed on the left to show that the dual is taken
on the indices $ab$  although it should be noted that the above equation
is equivalent to the same equation with the dual placed on the
right. It is clear that
one can define {\sl purely Weyl electric spacetimes} in an analogous fashion,
by replacing the Riemann tensor in (\ref{eq.el}) by the Weyl tensor.
Purely electric spacetimes were studied by Tr\"{u}mper \cite{trump} who showed
that
they are necessarily purely Weyl electric.
There is, however, no {\sl direct}
analogue  of Tr\"{u}mper's result for purely magnetic spacetimes, as
illustrated by an example discussed  by Arianrhod et al. \cite{arianrhod}.
A theorem concerning the relationship between purely magnetic and purely Weyl
magnetic spacetimes can, however, be formulated and will
be given  in section 3.
A spacetime which is
purely Weyl magnetic and purely Weyl electric necessarily has vanishing
Weyl tensor \cite{kramer}, but it is possible for a non-flat, non-vacuum
spacetime
to satisfy both conditions (\ref{eq.mag}) and (\ref{eq.el}) simultaneously
and this possibility  will be discussed in section 3.

\section{Purely Magnetic Vacuum Spacetimes}

Suppose that $(M,g)$ is a spacetime
and  at some $p\in M$ the following equation
holds for the Weyl tensor $C_{abcd}$ and a unit timelike vector $u^a$.
\begin{equation}
C_{abcd}u^bu^d=0\label{eq.weyl}
\end{equation}
The algebraic consequences of this condition on the Weyl tensor will
now be deduced.

The Petrov types of the Weyl
tensor correspond to  the possible
 Jordan canonical forms of the $3\times 3$
complex symmetric matrix $P_{\alpha\beta}$ defined by the following
equation  \cite{e_k} (Greek letters run from 1 to 3).
\begin{equation}
P_{\alpha\beta}=C_{\alpha 4\beta 4}+i\dual{C}_{\alpha 4\beta 4}
\label{petrov}
\end{equation}
In the above $C_{abcd}$ are the components of the Weyl tensor
at $p$ with respect to any orthonormal frame in which
the metric is given by $g=$diag$(1,1,1,-1)$ and,
since the left and right duals of the Weyl tensor are equal, one
can place the asterisk as above without ambiguity.
If there exists a solution $u^a$
of (\ref{eq.weyl}) then, in an orthonormal
 frame whose timelike member is $u^a$, $C_{\alpha 4\beta 4}=0$ and so
 the matrix
$P_{\alpha\beta}$ is a purely imaginary symmetric matrix, and hence
diagonalisable. This implies that the Petrov type of the  Weyl tensor
at $p$ is either $I$, $D$ or $O$ with
purely imaginary Petrov scalars. Conversely if $P_{\alpha\beta}$
assumes diagonal form with purely imaginary entries in some
orthonormal frame then from (\ref{petrov}) one has that
$C_{\alpha 4\beta 4}=0$ and hence the timelike member of this frame
satisfies (\ref{eq.weyl}).
 The vector  $u^a$ in (\ref{eq.weyl})  is the timelike
member of the canonical Petrov tetrad
and hence its uniqueness properties can be obtained from \cite{e_k}.
The algebraic information available from equation
(\ref{eq.weyl}) is contained in  the following theorem, which extends
results in  \cite{mcintosh,beem_h}.

\begin{theorem}

Suppose that $(M,g)$ is a spacetime and $p\in M$. It follows that
the Weyl tensor at $p$  satisfies (3) for a unit timelike vector field $u^a$
if and only if it is of Petrov type $I$, $D$ or $O$ with
purely imaginary Petrov scalars. In the  case of Petrov type $I$,
$u^a$ is unique up to sign  and in the case of Petrov type $D$,
$u^a$ is determined up to arbitrary boosts in the 2-space spanned by the
principal
null directions.
\end{theorem}

In the case of
a purely magnetic vacuum spacetime it can be shown using the Bianchi
identities that the Weyl tensor cannot be of  Petrov type $D$ over a non-empty
open set
\cite{mcintosh} (this result was also given by
 Hall~\cite{hall-petrov} in a slightly
different context).
If this fact is combined with Theorem 1 then one has

\begin{theorem}

Suppose that $(M,g)$ is a purely magnetic vacuum spacetime which
is non-flat in the sense that its Riemann tensor does not vanish over
an non-empty
 open set. The Weyl tensor is then of Petrov type $I$ almost everywhere
(i.e.~on an open dense subset of $M$) and the unit
 timelike vector $u^a$ is unique
up to sign.

\end{theorem}

The final result in this section will concern the kinematic
properties of the timelike congruence associated with a solution
$u^a$ of (\ref{eq.weyl}). The covariant derivative of a unit
timelike vector field $u_a$ can be decomposed in the following
well known way (see e.g.~\cite{kramer,e_k})
\begin{equation}
u_{a;b}=-\dot{u}_au_b+\omega_{ab}+\sigma_{ab}+{\sfrac{1}{3}}
\theta h_{ab}
\label{eq.kin}
\end{equation}
In equation (\ref{eq.kin})  $\dot{u}_a\equiv u_{a;b}u^b$ ;
$\omega_{ab}=\omega_{[ab]}$ ; $\sigma_{ab}=\sigma_{(ab)}$
 and $h_{ab}\equiv g_{ab}+u_au_b$ where square brackets denote
skew-symmetrisation, round brackets denote symmetrisation and
a semi-colon denotes a  covariant derivative. The quantities
$\omega_{ab}$, $\sigma_{ab}$ and $\theta$ are known  respectively
as the {\sl rotation}, {\sl shear} and {\sl expansion}
of the congruence. If the shear and rotation of a
timelike congruence in some spacetime vanish then the spacetime
must be purely Weyl electric \cite{trump}.  For purely
magnetic vacuum spacetimes the properties of the timelike
congruence associated with $u^a$  are controlled by the
following theorem.

\begin{theorem}

Let $(M,g)$ be a purely magnetic non-flat vacuum spacetime
with $u^a$ the (necessarily unique up to sign) unit timelike
vector field satisfying (\ref{eq.weyl}). It then follows that
the shear of the timelike congruence associated with $u^a$
cannot vanish over a non-empty open set
and if the rotation is identically zero  then the shear
necessarily assumes a diagonal form with respect to the canonical
Petrov tetrad.

\end{theorem}

{}From (\ref{eq.weyl}) it follows that $(C_{abcd}u^bu^d)^{;a}=0$
and if the vacuum Bianchi identities are used together with
the cyclic symmetry $C_{a[bcd]}=0$ then one obtains
\begin{equation}
 C_{abcd}u^d(2u^{b;c}-u^{c;b})=0
\end{equation}
Now if (\ref{eq.kin}) is used to substitute for $u_{a;b}$ then
it can be seen that
\begin{equation}
C_{abcd}u^d(3\omega^{bc}+\sigma^{bc})=0
\label{eq.twist}
\end{equation}
Attention will now be restricted to a generic point $p\in M$
where it may be assumed (by theorem 2) that the Weyl tensor
is of Petrov type $I$ and (by a theorem of Brans~\cite{brans})
that all the Petrov scalars are non-zero. Let
$x^a$, $y^a$, $z^a$, $u^a$ be a canonical Petrov tetrad
(unique up to reflections) and fix an orientation such that
this tetrad is positively oriented. A basis for the
space of 2-forms at $p$ is given by
$F_{ab}=u_{[a}x_{b]}$, $G_{ab}=u_{[a}y_{b]}$, $H_{ab}=u_{[a}z_{b]}$
and their duals. The canonical form for a Petrov
type $I$ Weyl tensor with purely imaginary Petrov scalars
is, with respect to the above basis~\cite{pirani}:
\begin{eqnarray}
C_{abcd}&=&\lambda_1(F_{ab}\dual{F}_{cd}+\dual{F}_{ab}F_{cd})+
           \lambda_2(G_{ab}\dual{G}_{cd}+\dual{G}_{ab}G_{cd}) \nonumber \\
        &  &\mbox{}+\lambda_3(H_{ab}\dual{H}_{cd}+\dual{H}_{ab}H_{cd})
\label{eq.typeI}
\end{eqnarray}
where $\lambda_1$, $\lambda_2$ and $\lambda_3$ are real and satisfy
$\lambda_1+\lambda_2+\lambda_3=0$. If equation (\ref{eq.typeI}) is
substituted into  (\ref{eq.twist}) then one obtains
the following three equations
\begin{eqnarray}
3(\lambda_3+\lambda_2)z_{[b}y_{c]}\omega^{bc}+
(\lambda_2-\lambda_3)z_{(b}y_{c)}\sigma^{bc}&=&0  \label{one}\\
3(\lambda_1+\lambda_3)x_{[b}z_{c]}\omega^{bc}+
(\lambda_3-\lambda_1)x_{(b}z_{c)}\sigma^{bc}&=&0  \label{two}\\
3(\lambda_2+\lambda_1)y_{[b}x_{c]}\omega^{bc}+
(\lambda_1-\lambda_2)y_{(b}x_{c)}\sigma^{bc}&=&0 \label{three}\\
 \nonumber
\end{eqnarray}

Since $p$ is a generic point then one cannot have $\lambda_i=\pm \lambda_j$
for $i\neq j$. Suppose the
 shear  $\sigma_{ab}$
is identically zero on some non-empty open subset $U\subset M$. Equations
(\ref{one}) to (\ref{three}) will then show that the rotation
vanishes almost everywhere on $U$ and hence is identically
zero on $U$ by continuity. It then follows \cite{trump} that
the spacetime is purely electric on $U$ and hence $C_{abcd}=0$
on $U$, contradicting the non-flat assumption. The shear is therefore
non-zero on an open dense subset of $M$.

If the rotation is identically zero on $M$ then equations
(\ref{one}) to (\ref{three}) show that $\sigma_{ab}$ takes the
form $\alpha x_ax_b+\beta y_ay_b+\gamma z_az_b$ for some
real functions $\alpha$, $\beta$ and $\gamma$.$\Box$

\section{Purely Magnetic non-vacuum Spacetimes}

In this section consideration will be given to the existence
and uniqueness of solutions to equation (1) for non-vacuum spacetimes.
Unfortunately it has not proved possible to give a complete solution to
the problem as is provided by theorems 1 and 2 in the vacuum
case but, nevertheless, some algebraic information is available
about the Riemann tensor of purely magnetic non-vacuum spacetimes.
The question of under what circumstances  equation~(\ref{eq.mag}) implies
 equation~(\ref{eq.weyl}) is answered by theorem~5
and the consequences of (\ref{eq.mag}) for the  type of matter field
present  are given by theorem~7.

The first point to be remarked upon is that the terminology
`purely magnetic' is a little misleading in the non-vacuum case.
The terminology arose due to the fact that one can decompose the Weyl
tensor in a formally similar fashion to the decomposition
of the Maxwell bivector into electric and magnetic
components. The electric and magnetic components of the Weyl tensor
 (with respect to a given observer $u^a$) are, respectively,
 the real and imaginary parts of the matrix $P_{\alpha\beta}$
defined by equation (\ref{petrov}), where $u^a$ is the timelike
 member of the orthonormal frame.
The Weyl tensor can then be described as `purely magnetic'
if its electric part vanishes, and vice-versa.
If a similar construction
were to be attempted for the (non-vacuum) Riemann tensor then it
would be possible for the corresponding $P_{\alpha\beta}$ to
vanish for a non-zero Riemann tensor. The Weyl tensor decomposition
relies on the property ${}^*C_{abcd}=C^*_{abcd}$, which is not enjoyed
by the Riemann tensor. The consequence of the preceding remarks
is that one can construct  spacetimes which are purely
electric {\sl and} purely magnetic and such spacetimes are characterised
by the existence of a timelike $u^a$ satisfying
\begin{equation}
R_{abcd}u^d=0
\label{eq.elmag}
\end{equation}
The above equation has been studied in the context of the algebraic
determination of the metric by the Riemann tensor~\cite{hall_mc}
and  the algebraic properties of a Riemann tensor satisfying
(\ref{eq.elmag}) are derived in~\cite{hall_kay}. It is remarked
that a spacetime which is 1+3 locally decomposable~\cite{hall_kay1},
where the three dimensional  factor  is spacelike, has timelike
solutions to (\ref{eq.elmag}).
 The following
theorem summarises the results contained in the references given.

\begin{theorem}
Suppose that $(M,g)$ is a spacetime and at some $p\in M$ there is a unit
timelike vector  $u^a$  satisfying equations (\ref{eq.mag})
and (\ref{eq.el}). Then $u^a$ satisfies (\ref{eq.elmag}) and
the following two possibilities occur if the Riemann tensor is non-zero
at $p$.
\begin{description}
\item[(i)] Equation (\ref{eq.elmag}) has a  unit timelike solution
which is unique
up to sign,  the Weyl tensor is of Petrov type $I$, $D$ or $O$ and
the Ricci tensor is diagonalisable.
\item[(ii)] Equation (\ref{eq.elmag}) admits 2 independent  timelike
solutions.
The Weyl tensor is of Petrov type $D$ and the Ricci tensor
is of Segre type $\{(11)(1,1)\}$. The solutions to (\ref{eq.elmag})
consist of all the timelike vectors in the 2-space spanned by the
principal null directions of the Weyl tensor.
\end{description}
In addition, if condition (i) holds over  a non-empty open set
then it follows by an application of the Bianchi identities
that the congruence associated with $u^a$ is geodesic.
\end{theorem}

The next result in this section will concern the relationship
between purely magnetic and purely Weyl magnetic spacetimes. As
has been observed by Arianrhod et al.~\cite{arianrhod},
condition (\ref{eq.mag}) does not in general imply that
condition (\ref{eq.weyl}) holds. However if one imposes restrictions
on the Ricci tensor then it is found that these two conditions are equivalent.

\begin{theorem}

Let $(M,g)$ be a spacetime, $p\in M$ and $u^a$ a unit timelike
vector at $p$.  It follows that any two of the following conditions at $p$
implies the third.
\begin{description}
\item[(i)] $C_{abcd}u^bu^d=0$
\item[(ii)] $R_{abcd}u^bu^d=0$
\item[(iii)] The Ricci tensor ($R_{ab}\equiv R^c{}_{acb}$)
takes the form $u_{(a}q_{b)}-u^cq_cg_{ab}$ for some vector $q_a$.
\end{description}

\end{theorem}
\paragraph{Proof}
The Riemann tensor admits the standard decomposition
(see eg~\cite{kramer})
\begin{equation}
R_{abcd}=C_{abcd}+E_{abcd}+{\sfrac{1}{6}}Rg_{a[c}g_{d]b}
\label{eq.decomp}
\end{equation}
where in the above the  following definitions have been used
\begin{eqnarray}
E_{abcd}&=&S_{a[c}g_{d]b}-S_{b[c}g_{d]a} \nonumber\\
S_{ab}&=&R_{ab}-{\sfrac{R}{4}}g_{ab}  \\
R&=&R^a{}_a \nonumber
\end{eqnarray}
If one contracts (\ref{eq.decomp}) with $u^bu^d$ and defines
$h_{ab}=g_{ab}+u_au_b$ then after some calculation one obtains
\begin{equation}
R_{abcd}u^bu^d=C_{abcd}u^bu^d-{\sfrac{1}{2}} h^b{}_ch^d{}_a[R_{bd}
-({\sfrac{R}{3}}+R_{ef}u^eu^f)g_{bd}]
\label{eq.result}
\end{equation}
If conditions (i) and (ii) hold then (\ref{eq.result}) shows that the
expression in the square brackets is of the form $u_{(a}q_{b)}$
for some $q_b$ and, noting that (ii) gives $R_{ab}u^au^b=0$,
a little rearrangement shows that $R_{ab}$ takes the form given in (iii).
If (iii) holds then equation (\ref{eq.result})   reduces to
$R_{abcd}u^bu^d=C_{abcd}u^bu^d$ and the rest of the theorem follows.$\Box$

The possible Segre types of a Ricci tensor of the form
given in condition (iii) of the above theorem are related to the
nature of the vector $q_a$. To calculate these Segre types
it will
prove more convenient to work with the tensor
$\tilde{R}_{ab}\equiv R_{ab}-(R/3)g_{ab}$
which clearly has the same Segre type as ${R}_{ab}$.
 It should first be noted that since  $\tilde{R}_{ab}= u_{(a}q_{b)}$
it necessarily has two spacelike eigenvectors with
eigenvalue zero spanning  the spacelike  2-space which is the  orthogonal
complement to the 2-space spanned by $q_a$ and $u_a$. If
$q_a\neq 0$ then this must be the only  spacelike
eigen-2-space, or else it is contained in a (necessarily spacelike)
eigen-3-space in which case the Segre type of $\tilde R_{ab}$ is
$\{(111),1\}$ with $u^a$ the unique timelike eigenvector
and $u^{[a}q^{b]}=0$. If $u_a$ is not
parallel to $q_a$  then any further eigenvectors  must lie in the timelike
 2-space spanned by $u_a$ and $q_a$. A short calculation
will show that the Segre type of $\tilde{R}_{ab}=u_{(a}q_{b)}$
is $\{(11)2\}$, $\{(11)z\bar{z}\}$ or
$\{(11)1,1\}$ according to whether $q_b$ is null, spacelike
or timelike respectively (where  $\bar{z}z$ denotes a pair
of conjugate complex eigenvectors).
 An examination of the appropriate
canonical forms given in \cite{hall_ricci} associated with these
Segre types will show that no further degeneracies are possible
(apart from $q_a$ parallel to $u_a$ or $q_a=0$). In addition
the expression $\tilde{R}_{ab}=u_{(a}q_{b)}$ uniquely determines
the vectors  $u_a$ and $q_a$ up to a possible scaling and interchange
of $u_a$ and $q_a$ (clearly the interchange is only possible
where $u_a$ and $q_a$ have the same nature.) The following
theorem summarises the foregoing discussion.

\begin{theorem}
Suppose that at some point of a spacetime the Ricci tensor $R_{ab}$
can be written in the form given in condition (iii) of the
previous theorem. Assuming $u_a$ and $q_a$ non-zero and
non-parallel then the Segre type of $R_{ab}$
is either $\{(11)1,1\}$, $\{(11)z\bar{z}\}$ or
$\{(11)2\}$ depending on whether $q_a$ is timelike, spacelike
or null respectively. No further degeneracies are possible.
The vectors $u_a$ and $q_a$ are unique up to scaling
and/or swapping. If $u_a$ and $q_a$
 are parallel then their direction
is unique and the Segre type of $R_{ab}$ is $\{(111),1\}$.
\end{theorem}

It should be emphasised that if the Ricci tensor is of one of the
Segre types given above then it does not necessarily take the form
given in condition (iii) of theorem 5, as additional restrictions
are placed on the eigenvalues by condition (iii).

The possible types of energy-momentum tensor
in a  purely magnetic spacetime are restricted by the fact that
equation (\ref{eq.mag}) can be contracted to give
the necessary condition
\begin{equation}
 R_{ab}u^au^b=0
\label{eq.ricci}
\end{equation}
It has been pointed out by Beem and Harris~\cite{beem_h} that
this equation cannot hold in a proper Einstein space
and it follows from a result of Penrose and Rindler~\cite[p328]{p_r}
that (\ref{eq.ricci}) cannot hold in a (null or non-null)
electromagnetic spacetime, with vanishing cosmological
constant. Suppose that the Ricci tensor
takes the algebraic form of a perfect fluid type
with flow vector $v^a$, pressure $p$ and energy density $\mu$.
The Ricci tensor can then be written as~\cite{kramer}
\begin{equation}
R_{ab}=(\mu+p)v_av_b+{\frac{1}{2}}(\mu-p)g_{ab}
\label{eq.fluid}
\end{equation}
Combining equations (\ref{eq.fluid}) and (\ref{eq.ricci})
gives that  the necessary and sufficient
condition for $R_{ab}$  to satisfy
(\ref{eq.ricci}) is   $2(v^au_a)^2=(\mu-p)/(\mu+p)$.
In particular if one wishes to have $u^a=v^a$
then the appropriate condition is $\mu+3p=0$ and in this case
it follows that $R_{ab}$ satisfies condition (iii) of theorem 5
and hence
the Weyl tensor $C_{abcd}$ satisfies $C_{abcd}u^bu^d=0$.
Conversely if $C_{abcd}u^bu^d=0=R_{abcd}u^bu^d$ when
the Ricci tensor is of the form (\ref{eq.fluid}) then theorems
5 and 6 show that $u^{[a}v^{b]}=0$.
The following theorem  summarises the possible  physical
types of purely magnetic spacetimes.

\begin{theorem}

Suppose that $(M,g)$ is a spacetime and that equation
(\ref{eq.mag}) holds at some $p\in M$. It then follows that the
Ricci tensor cannot be a non-zero multiple of the metric
or be of electromagnetic type at $p$.
If the Ricci tensor is of perfect fluid type with flow vector
$v^a$, energy density $\mu$ and pressure $p$  then one has
\begin{equation}
\mu=\left( {\frac{1+2(v_au^a)^2}{1-2(v_au^a)^2}} \right) p
\end{equation}
In this case if
 $u^{[a}v^{b]}=0$ if and only if  the Weyl tensor is of petrov type $I$, $D$ or
$O$ with purely imaginary Petrov scalars.

\end{theorem}

In the non-vacuum case the uniqueness of a solution $u^a$ to
(\ref{eq.mag}) is, in general,
 much harder to predict than in the vacuum case.
Some results in this respect have been given by Beem and Harris
\cite{beem_h}, for example if $u^a$, $v^a$ and some linear combination
of $u^a$ and $v^a$ satisfy (\ref{eq.mag}) then all linear combinations
of $u^a$ and $v^a$ will satisfy (\ref{eq.mag}). Various
generalisations of this result are given in \cite{beem_h}.

In some circumstances the Riemann tensor may be invariant
under certain Lorentz transformations of the tangent space
(at some point $p$)
and these transformations form the {\sl isotropy group}
of the Riemann tensor at $p$. If, under the
action of a member $A$ of the isotropy group,
$u^a$ is mapped to some other vector $v^a$ then $v^a$ must
also be a solution of (\ref{eq.mag}).
In the case where $R_{abcd}=C_{abcd}$ and a unit timelike non-trivial
solution $u^a$
of (\ref{eq.mag}) is  admitted then all other solutions
to this equation can be found by considering the orbit of $u^a$
under the isotropy group of the Riemann tensor (cf theorem 2). Unfortunately
such a method cannot be used in the general case as the following
example will show. If $x_a,y_a,z_a,v_a$ is an orthonormal
frame for the cotangent space at $p$ then define
\begin{equation}
\tilde{R}_{abcd}=x_{[a}y_{b]}y_{[c}z_{d]}+y_{[a}z_{b]}x_{[c}y_{d]}
\label{eq.isotropy}
\end{equation}
It is easily seen  that $\tilde{R}_{abcd}$ has all the algebraic symmetries
of a Riemann tensor. It may also be  verified that if $u_a$
is orthogonal to the 2-plane spanned by $x_a$ and $y_a$ or if $u_a$
is orthogonal to the 2-plane spanned by $y_a$ and $z_a$ then
$\tilde{R}_{abcd}u^bu^d=0$. However if one calculates
the Weyl tensor $\tilde{C}_{abcd}$ of
$\tilde{R}_{abcd}$, by noting that $\tilde{R}_{ab}{}^{ab}=0$
and hence $2\tilde{C}_{abcd}=\tilde{R}_{abcd}-{}^*\tilde{R}^*_{abcd}$ then
$\tilde{C}_{abcd}$ is seen to be of Petrov type I with
$v^a$ the timelike member of the canonical Petrov tetrad.
 Now the space of type
(0,4) tensors with Riemann symmetries decomposes under
the action of the Lorentz group into three irreducible subspaces
\cite{p_r} which correspond to the decomposition given by equation
(\ref{eq.decomp}). Consequently the isotropy of the Riemann
tensor is contained in that of the
Weyl tensor, which only consists of finite number of
 reflections in the case of
Petrov type I~\cite{p_r2}. The isotropy group of $\tilde{R}_{abcd}$
given by (\ref{eq.isotropy}) therefore  cannot generate  the infinite
number of  solutions
of (\ref{eq.mag}).

\section{Conclusions}

After having studied the algebraic properties of curvature
in purely magnetic spacetimes, one question that may come to  mind
is `are there any physically interesting examples?' In fact
there are very few known examples of purely magnetic spacetimes
of any description. One example due to  Misra et al.~\cite{misra}
 is discussed by McIntosh et al.~\cite{mcintosh} and the latter
suggest that there may be no purely magnetic non-flat vacuum
solutions. The reasons
why there are few purely magnetic spacetimes in the literature
are essentially given by theorems 2 and 7. The former
states that all vacuum purely magnetic solutions are of type I,
of which there are few solutions known (the static
type I solutions cannot be purely magnetic~\cite{arianrhod}),
  and the latter
implies that most physically interesting matter solutions
cannot be purely magnetic. Of course it is no easy matter
to determine whether or not a given metric has a
purely magnetic curvature tensor. There is, however, a method
given by McIntosh et al. for deciding whether the Weyl tensor
of a given metric admits solutions to (\ref{eq.weyl}),
which is suitable for implementation within a computer
algebra system.  It would be a fairly straightforward matter to
use such a computer system to check the literature
for purely Weyl  magnetic spacetimes.

Hawking and Ellis~\cite{h_e} have excluded purely
magnetic spacetimes in the conditions
of their singularity theorems and argue that such spacetimes
are unphysical. However it should be remembered that
their argument only applies when one wishes to model the entire
universe. There is no obvious physical reason why a purely magnetic
solution should not act as model for some portion of spacetime.
It is hoped that theorem 3 may be of some help
in the vacuum case as the restrictions it places on the derivative
of $u^a$ could conceivably be used to simplify the field equations.

\section*{Acknowledgements}

Useful comments were gratefully received
from Graham Hall and Petros Florides.
The author acknowledges the financial support
of the European Union under the Human   Capital and  Mobility
Programme.

\end{document}